# Learning Hyperedge Replacement Grammars for Graph Generation

Salvador Aguinaga, David Chiang, and Tim Weninger[*]

Department of Computer Science and Engineering
University of Notre Dame

February 23, 2018

## Abstract

The discovery and analysis of network patterns are central to the scientific enterprise. In the present work, we developed and evaluated a new approach that learns the building blocks of graphs that can be used to understand and generate new realistic graphs. Our key insight is that a graph's clique tree encodes robust and precise information. We show that a Hyperedge Replacement Grammar (HRG) can be extracted from the clique tree, and we develop a fixed-size graph generation algorithm that can be used to produce new graphs of a specified size. In experiments on large real-world graphs, we show that graphs generated from the HRG approach exhibit a diverse range of properties that are similar to those found in the original networks. In addition to graph properties like degree or eigenvector centrality, what a graph "looks like" ultimately depends on small details in local graph substructures that are difficult to define at a global level. We show that the HRG model can also preserve these local substructures when generating new graphs.

## 1 Introduction

Teasing out signatures of interactions buried in overwhelming volumes of information is one of the most basic challenges in scientific research. Understanding how information is organized and how it evolves can help us discover its fundamental underlying properties. Researchers do this when they investigate the relationships between diseases, cell functions, chemicals, or particles, and we all learn new concepts and solve problems by understanding the relationships between the various entities present in our everyday lives. These entities can be represented as networks, or graphs, in which local behaviors can be understood, but whose global view is highly complex.

Discovering and analyzing network patterns to extract useful and interesting patterns (building blocks) is critical to the advancement of many scientific fields. Indeed the most pivotal moments in the development of a scientific field are centered on discoveries about the structure of some phenomena [1]. For example, biologists have agreed that tree structures are useful when organizing the evolutionary history of life [4, 5], and sociologists find that triadic closure underlies community development [6, 7]. In other instances, the structural organization of the entities may resemble a ring, a clique, a star, a constellation, or any number of complex configurations.

Unfortunately, current graph mining research deals with small pre-defined patterns [8, 9] or frequently reoccurring patterns [10, 11, 12, 13], even though interesting and useful information may be hidden in unknown and non-frequent patterns. Principled strategies for extracting these complex patterns are needed to discover the precise mechanisms that govern network structure and growth. In-depth examination of this mechanism leads to a better understanding of graph patterns involved in structural, topological, and functional properties of complex systems. This is precisely the focus of the present work: to develop and evaluate techniques that learn the building blocks of real-world systems that, in aggregate, succinctly describe the observed interactions expressed in a network.

---

[*]Corresponding Author: tweninge@nd.edu

These networks exhibit a long and varied list of global properties, including heavy-tailed degree distributions [14], and interesting community structures [15] to name a few. Recent work has found that these global properties are products of a graph's local properties [17, 18]. In the present work, our goal is to learn the local structures that, in aggregate, help describe the interactions observed in the network and generalize to applications across a variety of fields like computer vision, computational biology, and graph compression.

The key insight for this task is that a network's *clique tree* encodes robust and precise information about the network. A *hyperedge replacement grammar* (HRG), extracted from the clique tree, contains graphical rewriting rules that can match and replace graph fragments similar to how a context-free grammar (CFG) rewrites characters in a string. These graph fragments represent a succinct, yet complete description of the building blocks of the network, and the rewriting rules of the HRG describe the instructions on how the graph is pieced together.

The HRG framework is divided into two steps: 1) graph model learning and 2) graph generation. After reviewing some of the theoretical foundations of clique trees and HRGs, we show how to extract an HRG from a graph. These graph rewriting rules can be applied randomly to generate larger and larger graphs. However, scientists typically have a specific size in mind, so we introduce a fixed-size graph generation algorithm that will apply HRG rules to generate a realistic graph of a user-specified size.

Finally, we present experimental results that compare the generated graphs with the original graphs. We show that these generated graphs exhibit a broad range of properties that are very similar to the properties of the original graphs and outperform existing graph models across a variety of graph comparison metrics.

## 2 Preliminaries

The new work in this paper begins where previous work [21, 22, 23, 24] left off. However, before we begin, some background knowledge is crucial to understand the key insights of our main contributions.

We begin with an arbitrary input *hypergraph* $H = (V, E)$, where $V$ is a finite set of vertices and $E \subseteq V^+$ is a set of *hyperedges*. A hyperedge $e \in E$ can connect one or more ordered vertices and is written $e = (v_1, v_2, \ldots, v_k)$. Common *graphs* (e.g., social networks, Web graphs, information networks) are a particular case of hypergraphs where each edge connects exactly two vertices. For convenience, all of the graphs in this paper will be *simple*, *connected* and *undirected*, although these restrictions are not vital. In the remainder of this section, we refer mainly to previous developments in clique trees and their relationship to hyperedge replacement grammars in order to support the claims made in sections 3 and 4.

### 2.1 Clique Trees

All graphs can be decomposed (though not uniquely) into a *clique tree*, also known as a tree decomposition, junction tree, join tree, intersection tree, or cluster graph. Within the pattern recognition community, clique trees are best known for their role in exact inference in probabilistic graphical models, so we introduce the preliminary work from a graphical modeling perspective; for an expanded introduction, we refer the reader to Chapters 9 and 10 of Koller and Friedman's textbook [25].

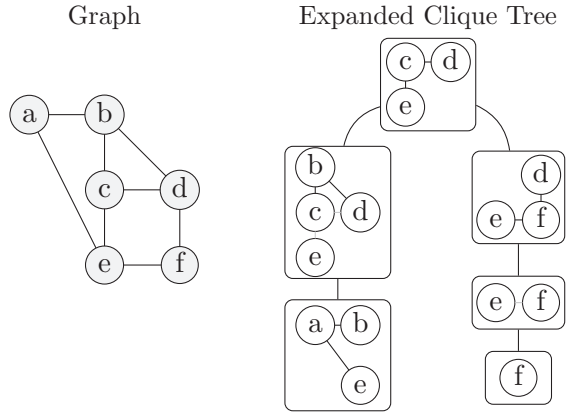

Figure 1: A graph and one possible minimal-width clique tree for it. Ghosted edges are not part of $E_\eta$; they are shown only for explanatory purposes.

**Definition 2.1** *A* clique tree *of a graph $H = (V, E)$ is a tree $CT$, each of whose nodes $\eta$ is labeled with a $V_\eta \subseteq V$ and $E_\eta \subseteq E$, such that the following properties hold:*

1. *Vertex Cover: For each $v \in V$, there is a vertex $\eta \in CT$ such that $v \in V_\eta$.*

2. *Edge Cover: For each hyperedge $e_i = \{v_1, \ldots, v_k\} \in E$ there is exactly one node $\eta \in CT$ such that $e \in E_\eta$. Moreover, $v_1, \ldots, v_k \in V_\eta$.*

3. *Running Intersection: For each $v \in V$, the set $\{\eta \in CT \mid v \in V_\eta\}$ is connected.*

**Definition 2.2** *The* width *of a clique tree is $\max(|V_\eta - 1|)$, and the* treewidth *of a graph $H$ is the minimal width of any clique tree of $H$.*

Unfortunately, finding the optimal elimination ordering and corresponding minimal-width clique tree is NP-Complete [26]. Fortunately, many reasonable approximations exist for general graphs: in this paper, we employ the commonly used maximum cardinality search (MCS) heuristic introduced by Tarjan and Yannikakis [27] to compute a clique tree with a reasonably-low, but not necessarily minimal, width.

Simply put, a clique tree of any graph (or any hypergraph) is a tree. Each of whose nodes we label with nodes and edges from the original graph, such that *vertex cover*, *edge cover* and the *running intersection* properties hold, and the "width" of the clique tree measures how tree-like the graph is. The reason for the interest in finding the clique tree of a graph is because many computationally difficult problems can be solved efficiently when the data is constrained to be a tree.

Figure 1 shows a graph and its minimal-width clique tree (showing $V_\eta$ for each node $\eta$). We label nodes with lowercase Latin letters. We will refer back to this graph and clique tree as a running example throughout this paper.

## 2.2 Hyperedge Replacement Grammar

The key insight for this task is that a network's clique tree encodes robust and precise information about the network. An HRG, extracted from the clique-tree, contains graphical rewriting rules that can match and replace graph fragments similar to how a context-free Grammar (CFG) rewrites characters in a string. These graph fragments represent a succinct, yet complete description of the building blocks of the network, and the rewriting rules of the HRG describe the instructions on how the graph is pieced together. For a thorough examination of HRGs, we refer the reader to the survey by Drewes *et al.* [28].

**Definition 2.3** *A* hyperedge replacement grammar *is a tuple* $G = \langle N, T, S, \mathcal{P} \rangle$, *where*

1. *$N$ is a finite set of nonterminal symbols. Each nonterminal $A$ has a nonnegative integer* rank, *which we write $|A|$.*
2. *$T$ is a finite set of terminal symbols.*
3. *$S \in N$ is a distinguished starting nonterminal, and $|S| = 0$.*
4. *$\mathcal{P}$ is a finite set of production rules $A \to R$, where*

   - *$A$, the left hand side (LHS), is a nonterminal symbol.*
   - *$R$, the right hand side (RHS), is a hypergraph whose edges are labeled by symbols from $T \cup N$. If an edge $e$ is labeled by a nonterminal $B$, we must have $|e| = |B|$.*
   - *Exactly $|A|$ vertices of $R$ are designated* external vertices *and numbered $1, \ldots, |A|$. The other vertices in $R$ are called* internal *vertices.*

When drawing HRG rules, we draw the LHS $A$ as a hyperedge labeled $A$ with arity $|A|$. We draw the RHS as a hypergraph, with external vertices drawn as solid black circles and the internal vertices as open white circles.

If an HRG rule has no nonterminal symbols in its RHS, we call it a *terminal rule*. If an HRG rule has exactly one nonterminal symbol in its RHS, we call it a *unary rule*.

**Definition 2.4** *Let $G$ be an HRG and $P = (A \to R)$ be a production rule of $G$. We define the relation $H' \Rightarrow H^*$ ($H^*$ is derived in one step from $H'$) as follows. $H'$ must have a hyperedge $e$ labeled $A$; let $v_1, \ldots, v_k$ be the vertices it connects. Let $u_1, \ldots, u_k$ be the external vertices of $R$. Then $H^*$ is the graph formed by removing $e$ from $H'$, making an isomorphic copy of $R$, and identifying $v_i$ with the copies of $u_i$ for each $i = 1, \ldots, k$.*

*Let $\Rightarrow^*$ be the reflexive, transitive closure of $\Rightarrow$. Then we say that $G$ generates a graph $H$ if there is a production $S \to R$ and $R \Rightarrow^* H$ and $H$ has no edges labeled with nonterminal symbols.*

In other words, a derivation starts with the symbol $S$, and we repeatedly choose a nonterminal $A$ and rewrite it using a production $A \to R$. The replacement hypergraph fragments $R$ can itself have other nonterminal hyperedges, so this process repeats until there are no more nonterminal hyperedges. The following sections illustrate these definitions more clearly.

## 3 Learning HRGs

The first step in learning an HRG from a graph is to compute a clique tree from the original graph. Then, this clique-tree directly induces an HRG, which we demonstrate in this section.

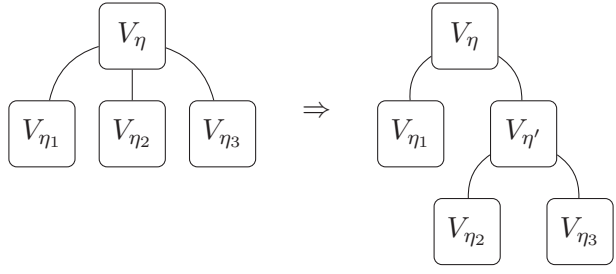

Figure 2: Binarization of a bag in a clique tree.

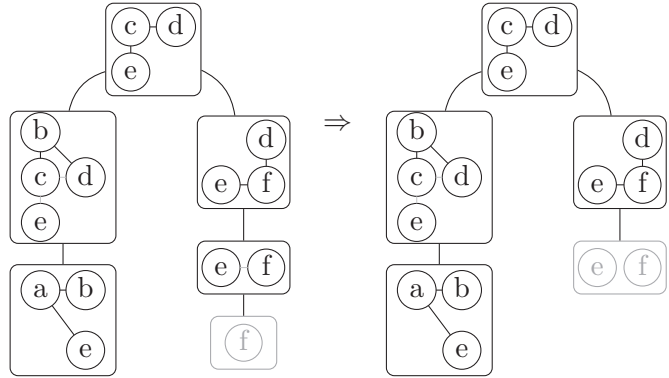

Figure 3: Pruning a clique tree to remove leaf nodes without internal vertices. Ghosted clique tree nodes show nodes that are pruned.

## 3.1 Binarization

Just as context-free string grammars are more convenient to parse if put into Chomsky normal form (CNF), we also assume, without loss of generality, that our HRG also follows CNF. This means that each rule's right-hand side has at most two nonterminals. By the HRG induction methods presented later in this section, each clique tree node $\eta$ yields an HRG rule, and the number of children of $\eta$ determines the number of nonterminals on the right-hand side of the resulting rule. Thus, it suffices for the clique tree to have a branching factor of at most two. Although the branching factor of a clique tree may be greater than two, it is always easy to binarize it.

There is more than one way to do this; we use the following scheme. Let $\eta$ be a clique tree node with children $\eta_1, \ldots, \eta_d$, where $d > 2$ (here $d$ corresponds to the number of children for a given parent node). These are labeled with bags $V_\eta, V_{\eta_1}, \ldots, V_{\eta_d}$, respectively. Make a copy of $\eta$; call it $\eta'$, and let $V_{\eta'} = V_\eta$. Let the children of $\eta$ be $\eta_1$ and $\eta'$, and let the children of $\eta'$ be $\eta_2, \ldots, \eta_r$. See Fig. 2 for an example. Then, if $\eta'$ has more than two children, apply this procedure recursively to $\eta'$.

It is easy to see that this procedure terminates and results in a clique tree whose nodes are at most binary-branching and still has the vertex cover, edge cover, and running intersection properties for $H$.

## 3.2 Clique Tree Pruning

Later we will introduce a dynamic programming algorithm for constructing graphs that require every leaf node of the clique tree to have at least one internal vertex. Clique tree algorithms, such as the MCS algorithm used in this paper, do no guarantee these conditions. Fortunately, we can just remove these leaf nodes from the clique tree.

The bottom-right clique tree node in Fig. 1 is such an example because f is an external vertex; that is, f exists in its parent. Because no internal vertices exist in this leaf node, it is removed from the clique tree. The clique tree node with vertices e and f is now a leaf, as illustrated in the left side of Fig. 3. Vertices e and f in the new leaf node are still both external vertices, so this clique tree node must also be removed creating a final clique tree illustrated in the right side of Fig. 3.

## 3.3 Clique Trees and HRGs

Here we show how to extract an HRG from the clique tree. Let $\eta$ be an interior node of the clique tree $CT$, let $\eta'$ be its parent, and let $\eta_1, \ldots, \eta_m$ be its children. Node $\eta$ corresponds to an HRG production rule $A \to R$ as follows. First, $|A| = |V_{\eta'} \cap V_\eta|$. Then, $R$ is formed by:

- Adding an isomorphic copy of the vertices in $V_\eta$ and the edges in $E_\eta$
- Marking the (copies of) vertices in $V_{\eta'} \cap V_\eta$ as external vertices
- Adding, for each $\eta_i$, a nonterminal hyperedge connecting the (copies of) vertices in $V_\eta \cap V_{\eta_i}$.

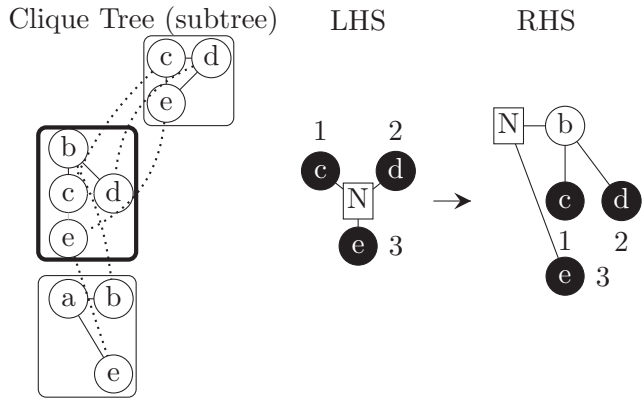

Figure 4: Example of hyperedge replacement grammar rule creation from an interior vertex of the clique tree. Note that lowercase letters inside vertices are for explanatory purposes only; only the numeric labels outside external vertices are actually part of the rule.

Figure 4 shows an example of the creation of an HRG rule. In this example, we focus on the middle clique-tree node $V_\eta = \{b, c, d, e\}$, outlined in bold.

We choose nonterminal symbol N for the LHS, which must have rank 3 because $\eta$ has 3 vertices in common with its parent. The RHS is a graph whose vertices are (copies of) $V_\eta = \{b, c, d, e\}$. Vertices c, d and e are marked external (and numbered 1, 2, and 3, arbitrarily) because they also appear in the parent node. The terminal edges are $E_\eta = \{(b, c), (b, d)\}$. There is only one child of $\eta$, and the nodes they have in common are b and e, so there is one nonterminal hyperedge connecting b and e.

Next we deal with the special cases of the root and leaves.

**Root Node.** If $\eta$ is the root node, then it does not have any parent cliques, but may still have one or more children. Because $\eta$ has no parent, the corresponding rule has a LHS with rank 0 and a RHS with no external vertices. In this case, we use the start nonterminal symbol $S$ as the LHS, as shown in Fig. 5.

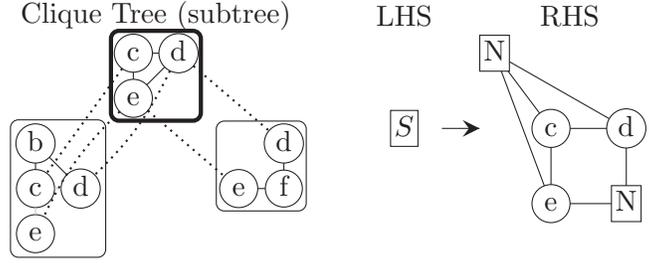

Figure 5: Example of hyperedge replacement grammar rule creation from the root node of the clique tree.

The RHS is computed in the same way as the interior node case. For the example in Fig. 5, the RHS has vertices that are copies of c, d, and e. In addition, the RHS has two terminal hyperedges, $E_\eta = \{(c, d), (c, e)\}$. The root node has two children, so there are two nonterminal hyperedges on the RHS. The right child has two vertices in common with $\eta$, namely, d and e; so the corresponding vertices in the RHS are attached by a 2-ary nonterminal hyperedge. The left child has three vertices in common with $\eta$, namely, c, d, and e, so the corresponding vertices in the RHS are attached by a 3-ary nonterminal hyperedge.

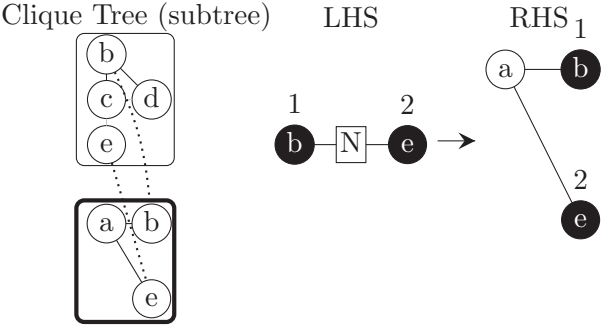

Figure 6: Example of hyperedge replacement grammar rule creation from a leaf vertex of the clique tree.

**Leaf Node.** If $\eta$ is a leaf node, then the LHS is calculated the same as in the interior node case. Again we return to the running example in Fig. 6. Here, we focus on the leaf node $\{a, b, e\}$, outlined in bold. The LHS has rank 2, because $\eta$ has two vertices in common with its parent.

The RHS is computed in the same way as the interior node case, except no new nonterminal hyperedges are added to the RHS. The vertices of the RHS are (copies of) the nodes in $\eta$, namely, a, b, and e. Vertices b and e are external because they also appear in the parent clique. This RHS has two terminal hyperedges, $E_\eta = \{(a, b), (a, e)\}$. Because the leaf clique has no children, it cannot produce any nonterminal hyperedges on the RHS; therefore this rule is a terminal rule.

### 3.4 Top-Down HRG Rule Induction

We induce production rules from the clique tree by applying the above extraction method top down. Because trees are acyclic, the traversal order does not matter, yet there are some interesting observations we can make about traversals of moderately sized graphs. First, exactly one HRG rule will have the special starting nonterminal $S$ on its LHS; no mention of $S$ will ever appear in any RHS. Similarly, the number of terminal rules is equal to the number of leaf nodes in the clique tree.

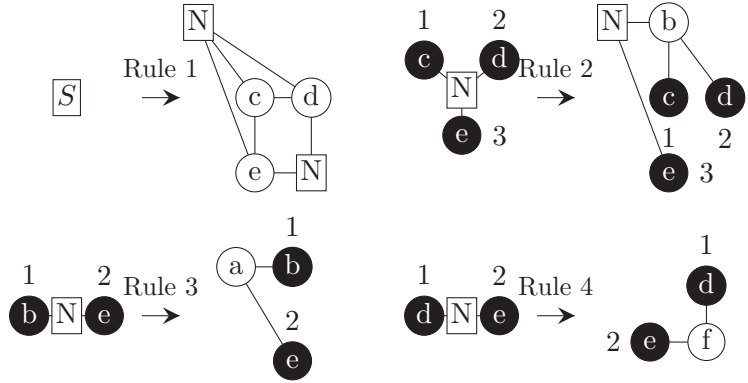

Figure 7: Complete set of production rules extracted from the example clique tree. Note that lowercase letters inside vertices are for explanatory purposes only; only the numeric labels outside external vertices are part of the rule.

Larger graphs will typically produce larger clique trees, especially sparse graphs because they are more likely to have a greater number of small maximal cliques. These larger clique trees will produce a large number of HRG rules, one for each clique in the clique tree. Although it is possible to keep track of each rule and its traversal order, we find, and will later show in the experiments section, that the same rules often repeat many times.

Figure 7 shows the 4 rules that are induced from the clique tree illustrated in Fig. 1 and used in the running example throughout this section.

## 4 Graph Generation

In this section, we show how to use the HRG extracted from the original graph $H$ (as described in the previous section) to generate a new graph $H^*$. Ideally, $H^*$ will be similar to or have features that are akin or analogous to the original graph $H$. We present two generation algorithms. The first generates random graphs with similar characteristics to the original graph. The second is like it but generates random graphs that have a specified number of nodes.

### 4.1 Stochastic Generation

There are many cases in which we prefer to create very large graphs in an efficient manner that still exhibit the local and global properties of some given example graph. Here we describe a simple stochastic hypergraph generator that applies rules from the extracted HRG to efficiently create such graphs.

In larger HRGs we usually find many $A \to R$ production rules that are identical. We chose to consider rules that are identical *modulo* a permutation of their external vertices to be equivalent as well. We can merge these duplicates by matching rule-signatures in a dictionary and keep a count of the number of times that each distinct rule has been seen. For example, if there were some extra Rule #5 in Fig. 7 that was identical to, say, Rule #3, then we would simply note that we saw Rule #3 two times.

To generate random graphs from a probabilistic HRG (PHRG), we start with the special starting nonterminal $H' = S$. From this point, $H^*$ can be generated as follows: (1) Pick any nonterminal $A$ in $H'$; (2) Find the set of rules $(A \to R)$ associated with LHS $A$; (3) Randomly choose one of these rules with probability proportional to its count; (4) Choose an ordering of its external vertices with uniform probability; (5) Replace $A$ in $H'$ with $R$ to create $H^*$; (5) Replace $H'$ with $H^*$ and repeat until there are no more nonterminal edges.

## 4.2 Fixed-Size Generation

A problem we find with the stochastic generation procedure is that, although the generated graphs have the same mean size as the original graph, the variance is much too high to be useful. So we want to sample only graphs whose size is the same as the original graph's, or some other user-specified size. Naively, we can do this using rejection sampling: sample a graph, and if the size is not right, reject the sample and try again. However, this would be quite slow. Our implementation uses a dynamic programming approach to sample a graph with specified size, while using quadratic time and linear space, or approximately while using linear time and space.

More formally, the learned PHRG defines a probability distribution over graphs, $P(H^*)$. But rather than sampling from $P(H^*)$, we want to sample from $P(H^* \mid |H^*| = n)$, where $n$ is the desired size.

Here, the stochastic generation sampling procedure is modified to rule out all graphs of the wrong size, as follows. Define a *sized* nonterminal $X^{(\ell)}$ to be a nonterminal $X$ together with a size $\ell > 0$. If $n$ is the desired final size, we start with $S^{(n)}$, and repeatedly:

1. Choose an arbitrary edge labeled with a sized nonterminal (call it $X^{(\ell)}$).

2. Choose a rule from among all rules with LHS $X$.

3. Choose sizes for all the nonterminals in the rule's RHS such that the total size of the RHS is $\ell$.

4. Choose an ordering of the external vertices of the rule's RHS, with uniform probability.

5. Apply the rule.

A complication arises when choosing the rule and the RHS nonterminal sizes (steps 2 and 3) because the weights of these choices no longer form a probability distribution. Removing graphs with the wrong size causes the probability distribution over graphs to sum to less than one, and it must be renormalized [29]. To do this, we precompute a table of *inside probabilities* $\alpha[X, \ell]$ for $\ell = 1, \ldots, n$, each of which is the total weight of derivations starting with $X$ and yielding a (sub)graph of size exactly $\ell$. These are computed using dynamic programming, as shown in Algorithm 1.

If $X \to R$ is a HRG rule, define size($R$) to be the increase in the size of a graph upon application of rule $(X \to R)$. If size is measured in vertices, then size($R$) is the number of *internal* vertices in $R$.

Rules that are unary and have zero size require some special care because they do not change the size of the graph. If there is a unary size-zero rule $X \to Y$, we need to ensure that $\alpha[Y, \ell]$ is computed before $\alpha[X, \ell]$, or else the latter will be incorrect. Thus, in Algorithm 1, we start by forming a weighted directed graph $U$ whose nodes are all the nonterminals in $N$, and for every unary rule $X \xrightarrow{p} Y$, there is an edge from $X$ to $Y$ with weight $p$. We perform a topological sort on $U$, and the loop over nonterminals $X \in N$ is done in reverse topological order.

However, if $U$ has a cycle, then no such ordering exists. The cycle could apply an unbounded number of times, and we need to sum over all possibilities. Algorithm 2 handles this more general

**Algorithm 1:** Compute inside probabilities (no cycles of size-zero unary rules)

    compute digraph $U$ of unary size-zero rules;
    topologically sort $U$;
    assert ($U$ is acyclic);
    **for** $\ell \leftarrow 1, \ldots, n$ **do**
        **for** $X \in N$ *in reverse topological order* **do**
            **for** *rules* $X \xrightarrow{p} R$ **do**
                $\ell' = \ell - \text{size}(R)$;
                **if** *$R$ has no nonterminals and $\ell' = 0$* **then**
                      $\alpha[X, \ell] \mathrel{+}= p$;
                **else if** *$R$ has nonterminal $Y$* **then**
                      $\alpha[X, \ell] \mathrel{+}= p \times \alpha[Y, \ell']$;
                **else if** *$R$ has nonterminals $Y$ and $Z$* **then**
                      **for** $k \leftarrow 1, \ldots, \ell' - 1$ **do**
                          $\alpha[X, \ell] \mathrel{+}= p \times \alpha[Y, k] \times \alpha[Z, \ell' - k]$;

case [30]. We precompute the strongly connected components of $U$, for example, using Tarjan's algorithm, and for each component $C$, we form the weighted adjacency matrix of $C$; call this $U_C$. The matrix $U_C^* = \sum_{i=0}^{\infty} U_C^i = (I - U_C)^{-1}$ gives the total weight of all chains of unary rules within $C$. So, after computing all the $\alpha[X, \ell]$ for $X \in C$, we apply the unary rules by treating the $\alpha[X, \ell]$ (for $X \in C$) as a vector and left-multiplying it by $U_C^*$. Some tricks are needed for numerical stability; for details, please see the released source code at https://github.com/nddsg/PHRG/.

In principle, a similar problem could arise with binary rules. Consider a rule $X \to R$ where $R$ is zero-size and has two nonterminals, $Y$ and $Z$. If $\alpha[Y, 0] > 0$, then $\alpha[X, \ell]$ is defined in terms of $\alpha[Y, \ell]$, which could lead to a circularity. Fortunately, we can avoid such situations easily. Recall that after clique tree pruning (Sec. 3.2), every leaf of the clique tree has at least one internal vertex. In terms of HRG rules, this means that if $R$ has no nonterminals, then $\text{size}(R) > 0$. Therefore, we have $\alpha[X, 0] = 0$ for all $X$, and no problem arises.

Once we have computed $\alpha$, we can easily sample a graph of size $n$ using Algorithm 3. Initially, we start with the sized start nonterminal $S^{(n)}$. Then, we repeatedly choose an edge labeled with a sized nonterminal $X^{(\ell)}$, use the table $\alpha$ of inside probabilities to recompute the weight of all the rewriting choices quickly, sample one of them, and apply it.

### 4.3 Pruning inside probabilities

The slowest step in the above method is the precomputation of inside probabilities (Alg. 2), which is quadratic in the number of vertices. To speed up this step up, we observe that randomly generated graphs tend to be highly unbalanced in the sense that if a rule has two nonterminal symbols, one is usually much larger than the other (see Figure 8). This is related to the fact, familiar with the study of algorithms, that random binary search trees tend to be highly unbalanced [31].

Therefore, we can modify Algorithm 2 to consider only splits where at most (say) 1000 nodes go to one nonterminal and the rest of the nodes go the other. This makes the algorithm asymptotically linear.

**Algorithm 2:** Compute inside probabilities (general)

    compute weighted digraph $U$ of unary size-zero rules;
    find strongly connected components (scc's) of $U$;
    compute $U_C^*$ for each scc $C$;
    **for** $\ell \leftarrow 1, \ldots, n$ **do**
        **for** *scc's $C$ in reverse topological order* **do**
            **for** $X \in C$ **do**
                **for** *rules $X \xrightarrow{p} R$* **do**
                      $\ell' = \ell - \text{size}(R)$;
                      **if** *$R$ has no nonterminals and $\ell' = 0$* **then**
                          $\alpha[X, \ell] \mathrel{+}= p$;
                      **else if** *$R$ has nonterminal $Y$ and $\ell' < \ell$* **then**
                          $\alpha[X, \ell] \mathrel{+}= p \times \alpha[Y, \ell']$;
                      **else if** *$R$ has nonterminals $Y$ and $Z$* **then**
                          **for** $k \leftarrow 1, \ldots, \ell' - 1$ **do**
                              $\alpha[X, \ell] \mathrel{+}= p \times \alpha[Y, k] \times \alpha[Z, \ell' - k]$;
            **for** $X \in C$ **do**
                $\alpha[X, \ell] = \sum_{Y \in C} [U_C^*]_{XY} \times \alpha[Y, \ell]$;

## 5 Experiments

Here we show that HRGs contain rules that succinctly represent the global and local structure of the original graph. In this section, we compare our approach against some of the state-of-the-art graph generators. We consider the properties that characterize some real-world networks and compare the distribution of graphs generated using Kronecker Graphs, the Exponential Random Graph, Chung-Lu Graphs, and the graphs produced by the probabilistic hyperedge replacement graph grammar.

Like HRGs, the Kronecker and Exponential Random Graph Models learn parameters that can be used to approximately recreate the original graph $H$ or a graph of some other size such that the probabilistically generated graph holds many of the same properties as the original graph. The Chung-Lu graph model relies on node degree sequences to yield graphs that maintain this distribution. The probabilistically generated graphs are likely not isomorphic to the original graph. We can, however, still judge how closely the probabilistically generated graph resembles the original graph by comparing several of their properties.

### 5.1 real-world Datasets

To get a holistic and varied view of the strengths and weaknesses of HRGs in comparison to the other leading graph generation models, we consider real-world networks that exhibit properties that are both common to many networks across different fields, but also have certain distinctive properties.

The six real-world networks considered in this paper are described in Table 1. The networks vary in their number of vertices and edges as indicated, but also vary in clustering coefficient, diameter, degree distribution and many other graph properties. Specifically, Karate Club graph is

**Algorithm 3:** Generate a graph with $n$ nodes

$H \leftarrow S^{(n)}$;
**while** $H$ contains a nonterminal $X^{(\ell)}$ **do**
    **for** all rules $X \xrightarrow{p} R$ **do**
        $\ell' = \ell - \text{size}(R)$;
        **if** $R$ has no nonterminals and $\ell' = 0$ **then**
            $\text{weight}[R] = p$;
        **else if** $R$ has nonterminal $Y$ **then**
            $R' = R\{Y \mapsto Y^{(\ell')}\}$;
            $\text{weight}[R'] = p \times \alpha[Y, \ell']$;
        **else if** $R$ has nonterminals $Y$ and $Z$ **then**
            **for** $k \leftarrow 1, \ldots, \ell' - 1$ **do**
                $R' = R\{Y \mapsto Y^{(k)}, Z \mapsto Z^{(\ell'-k)}\}$;
                $\text{weight}[R'] = p \times \alpha[Y, k] \times \alpha[Z, \ell' - k]$;
    let $P(R) = \text{weight}[R] / \sum_{R'} \text{weight}[R']$;
    sample sized RHS $R$ from $P(R)$;
    choose ordering of the external vertices of $R$;
    $H \leftarrow H\{X^{(\ell)} \mapsto R\}$;

Table 1: Experimental Dataset

| Dataset Name | Nodes | Edges |
|---:|---:|---:|
| Karate Club | 34 | 78 |
| Proteins (metabolic) | 1,870 | 2,277 |
| arXiv GR-QC | 5,242 | 14,496 |
| Internet Routers | 6,474 | 13,895 |
| Enron Emails | 36,692 | 183,831 |
| DBLP | 317,080 | 1,049,866 |
| Amazon | 400,727 | 2,349,869 |
| Flickr | 105,938 | 2,316,948 |

a network of interactions between members of a karate club; the Protein network is a protein-protein interaction network of *S. cerevisiae* yeast; the Enron graph is the email correspondence graph of the now defunct Enron corporation; the arXiv GR-QC graph is the co-authorship graph extracted from the General Relativity and Quantum Cosmology section of arXiv; the Internet router graph is created from traffic flows through Internet peers; DBLP is the co-authorship graph from the DBLP dataset; Amazon is the co-purchasing network from March 12, 2003; and, finally, Flickr is a network created from photos taken at the same location.

In the following experiments, we use the larger networks (arXiv, Routers, Enron, DBLP, Amazon, Flickr) for network generation and the smaller networks (Karate, Protein) for a special graph extrapolation task. Datasets were downloaded from the SNAP and KONECT dataset repositories.

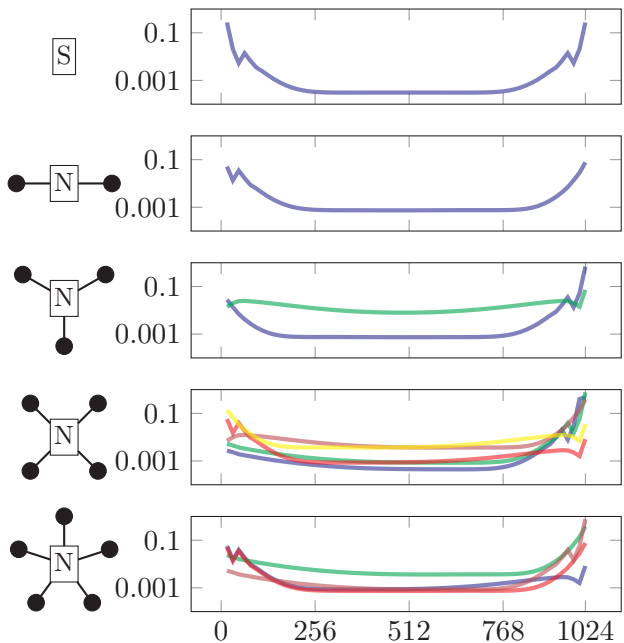

Figure 8: When an HRG rule has two nonterminal symbols, one is overwhelmingly likely to be much larger than the other. This plot shows, for various grammar rules (one LHS per row, one RHS per colored line), the probability (log scale) of apportioning 1024 nodes between two nonterminal symbols. This plot is best viewed in color.

## 5.2 Methodology

We compare several different graph properties from the four classes of graph generators (fixed-size HRG, Kronecker, Chung-Lu and exponential random graph (ERGM) models) to the original graph $H$. Other models, such as the Erdős-Rényi random graph model, the Watts-Strogatz small world model, the Barabási-Albert generator, etc. are not compared here because Kronecker, Chung-Lu and ERGM have been shown to outperform these earlier models when matching network properties in empirical networks.

Kronecker graphs operate by learning an initiator matrix and then performing a recursive multiplication of that initiator matrix to create an adjacency matrix of the approximate graph. In our case, we use KronFit [32] with default parameters to learn a $2 \times 2$ initiator matrix and then use the recursive Kronecker product to generate the graph. Unfortunately, the Kronecker product only creates graphs where the number of nodes is a power of 2, i.e., $2^x$, where we chose $x = 15$, $x = 12$, $x = 13$, and $x = 18$ for Enron, ArXiv, Routers and DBLP graphs respectively to match the number of nodes as close as possible.

The Chung-Lu Graph Model takes, as input, a degree distribution and generates a new graph of the similar degree distribution and size [33].

Exponential Random Graph Models are a class of probabilistic models. Their usefulness lies in that they directly describe several structural features of a graph [34]. We used default parameters in R's ERGM package [35] to generate graph models for comparison. In addition to the problem of model degeneracy, ERGMs do not scale well to large graphs. As a result, DBLP, Enron, Amazon, and Flickr could not be modelled due to their size, and the arXiv graph always resulted in a degenerate model. Therefore ERGM results are omitted from this section.

The main strength of HRG is to learn the patterns and rules that generate a large graph from only a few small subgraph-samples of the original graph. So, in all experiments, we make $k$ random samples of size $s$ node-induced subgraphs by a breadth first traversal starting from a random node in the graph [36]. By default we set $k = 4$ and $s = 500$ empirically. We then compute tree decompositions from the $k$ samples, learn HRGs $G_1, G_2, \ldots, G_k$, and combine them to create a single grammar $G = \bigcup_i G_i$.

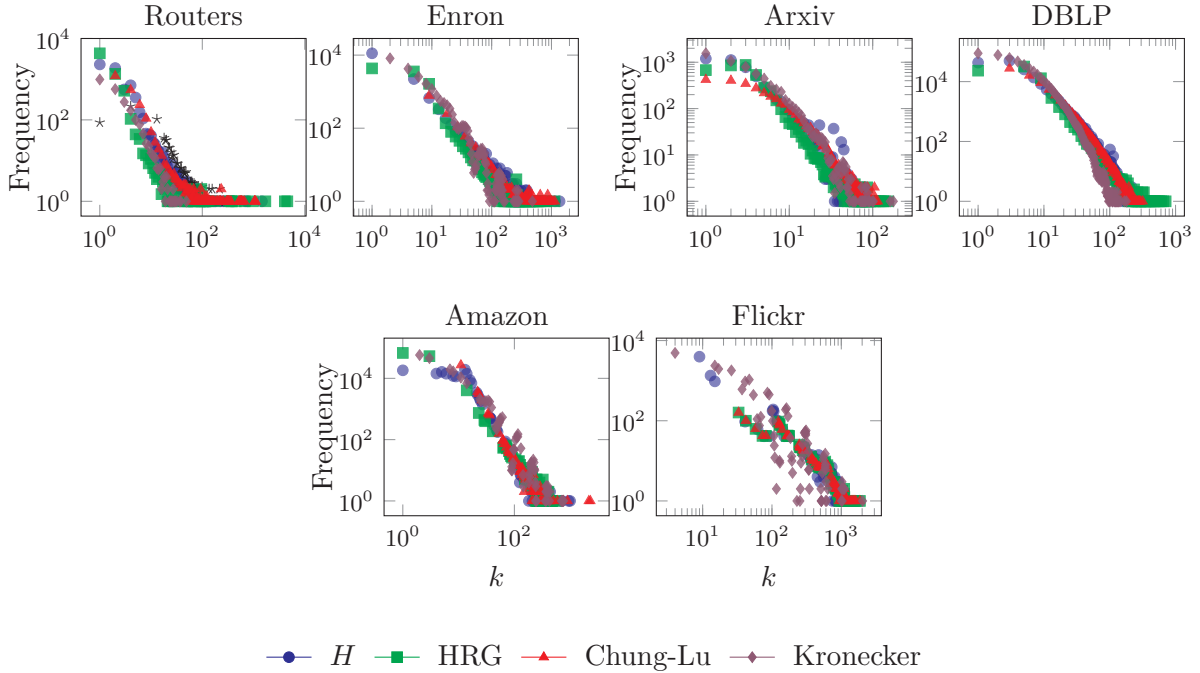

Figure 9: Degree Distribution. Dataset graphs exhibit a power law degree distribution that is well captured by existing graph generators as well as HRG.

Unless otherwise noted, we generate 20 graphs each for the HRG, Chung-Lu, and Kronecker models and plot the mean values in the results section. We did compute the confidence intervals for each of the models but omitted them from the graphs for clarity. In general, the confidence intervals were small for HRG, Kronecker, and Chung-Lu.

## 5.3 Graph Generation Results

Here we compare and contrast the results of approximate graphs generated from the HRG, Kronecker, and Chung-Lu models. Before presenting each result, we briefly introduce the graph properties that we used to compare the similarity between the real networks and their approximate counterparts. Although many properties have been discovered and detailed in related literature, we focus on five of the principal properties from which most others can be derived.

### 5.3.1 Global Measures

A key goal of graph modelling to preserve certain network properties of the original graph (i.e., $H$ as introduced in 2). Graphs generated using HRG, Kronecker, or Chung-Lu are analyzed by studying their fundamental network properties to assess how successful the model performs in generating graphs from parameters and production rules learned from the input graph. First, we look at the degree distribution, eigenvector centrality, local clustering coefficient, hop plot, and assortative mixing characteristics, and draw conclusions on these results.

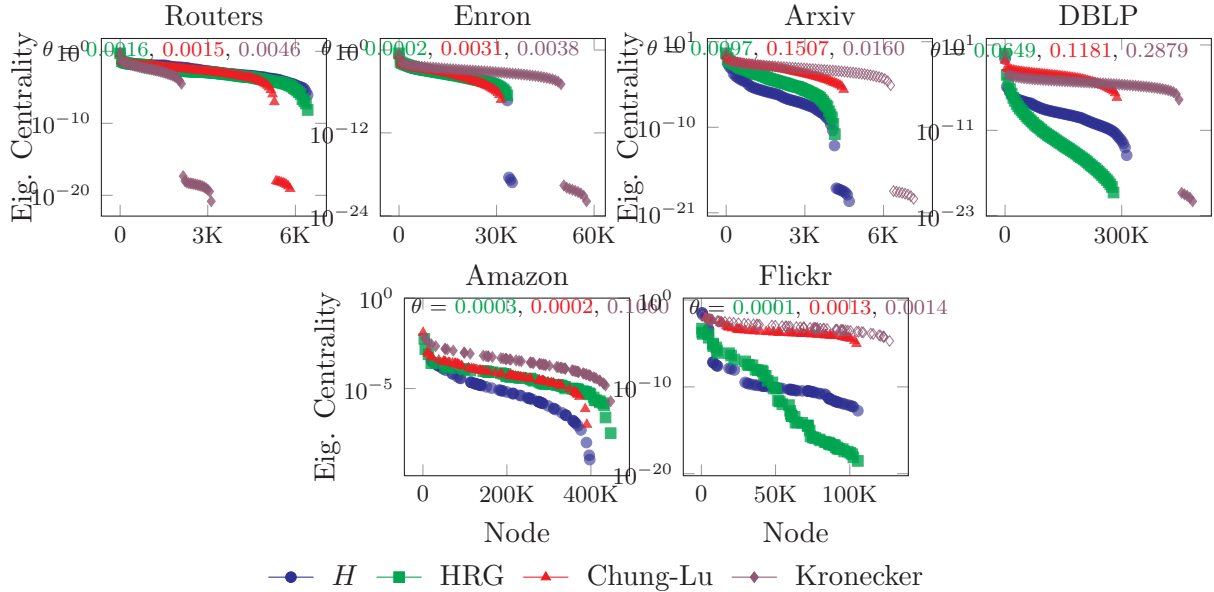

Figure 10: Eigenvector Centrality. Nodes are ordered by their eigenvector-values along the x-axis. Cosine distance between the original graph and HRG, Chung-Lu and Kronecker models are shown at the top of each plot where lower is better. In terms of cosine distance, the eigenvector of HRG is consistently closest to that of the original graph.

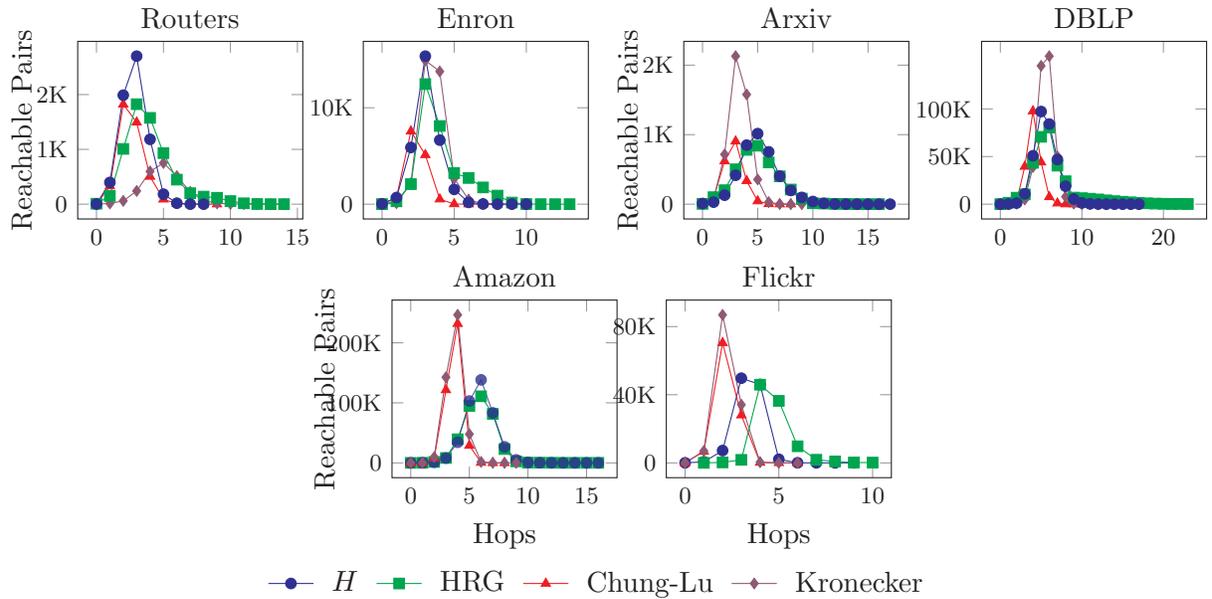

Figure 11: Hop Plot. Number of vertex pairs that are reachable within $x$-hops. HRG closely and consistently resembles the hop plot curves of the original graph.

**Degree Distribution.**   The degree distribution of a graph is the distribution of the number of edges connecting to a particular vertex. Figure 9 shows the results of the degree distribution property on the six real-world graphs. Recall that the graph results plotted here and throughout the results section are the mean averages of 20 generated graphs. Each of the generated graphs is slightly different from the original graphs in their own way. As expected, we find that the power law degree distribution is captured by existing graph generators as well as the HRG model.

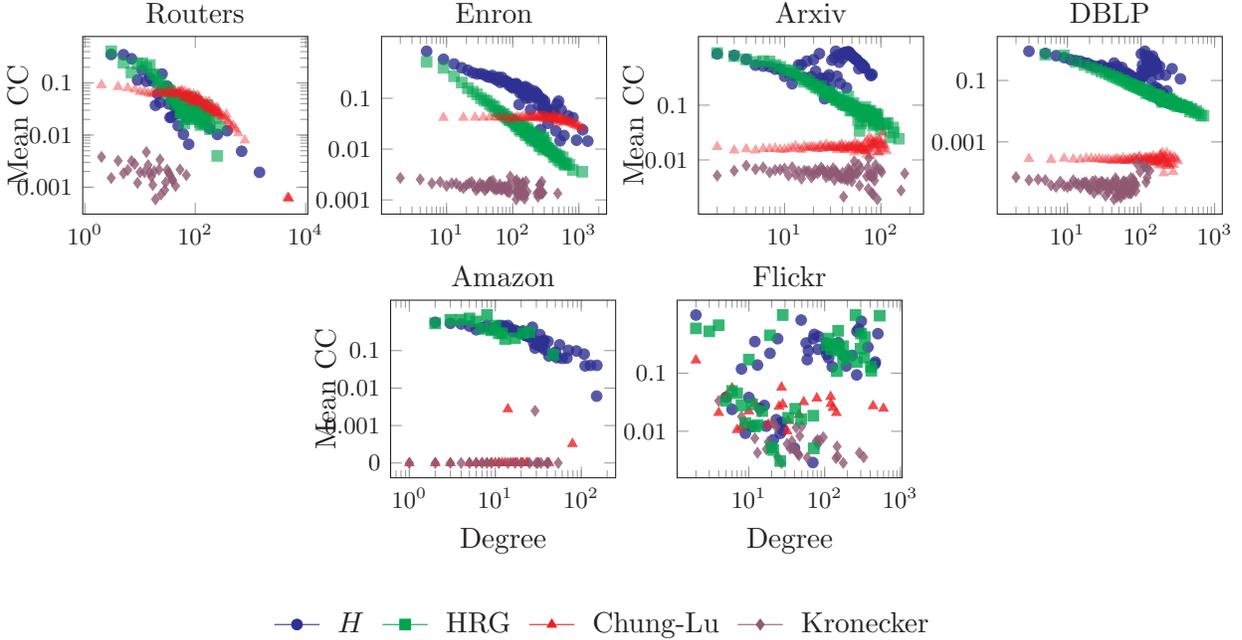

Figure 12: Mean Clustering Coefficient by Node Degree. HRG closely and consistently resembles the clustering coefficients of the original graph.

**Eigenvector Centrality.** The principal eigenvector is often associated with the centrality or "value" of each vertex in the network, where high values indicate an important or central vertex and lower values indicate the opposite. A skewed distribution points to a relatively few "celebrity" vertices and many common nodes.

The principal eigenvector value for each vertex is also closely associated with the PageRank and degree value for each node. Figure 10 shows the eigenvector scores for each node ranked highest to lowest in each of the six real-world graphs. Because the x-axis represents individual nodes, Fig. 10 also shows the size difference among the generated graphs. HRG performs consistently well across all graphs, but the log scaling on the y-axis makes this plot difficult to discern. To more concretely compare the eigenvectors, the pairwise cosine distance between eigenvector centrality of $H$ and the mean eigenvector centrality of each model's generated graphs appear at the top of each plot in order. HRG consistently has the lowest cosine distance followed by Chung-Lu and Kronecker.

**Hop Plot.** The hop-plot of a graph shows the number of vertex-pairs that are reachable within $x$ hops. The hop-plot, therefore, is another way to view how quickly a vertex's neighborhood grows as the number of hops increases. As in related work [16] we generate a hop-plot by picking 50 random nodes and performing a complete breadth first traversal over each graph. Figure 11 demonstrates that HRG graphs produce hop-plots that are remarkably similar to the original graph.

**Mean Clustering Coefficients.** A vertex's clustering coefficient is a measure of how well connected its neighbors are [40]. For each vertex in the graph, its clustering coefficient is the ratio of the number of edges in its ego-network (i.e., local neighborhood) to the total number of possible edges that could exist if the vertex's neighborhood was a clique. Calculating the clustering coefficient for each node is a computationally difficult task and difficult plot aesthetically, so we sampled 100 nodes from the graph randomly. Figure 12 shows the average clustering coefficients for the sampled nodes as a function of its degree in the graph. Like the results from Seshadhri et al., we find that the Kronecker and Chung-Lu models perform poorly at this task [15].

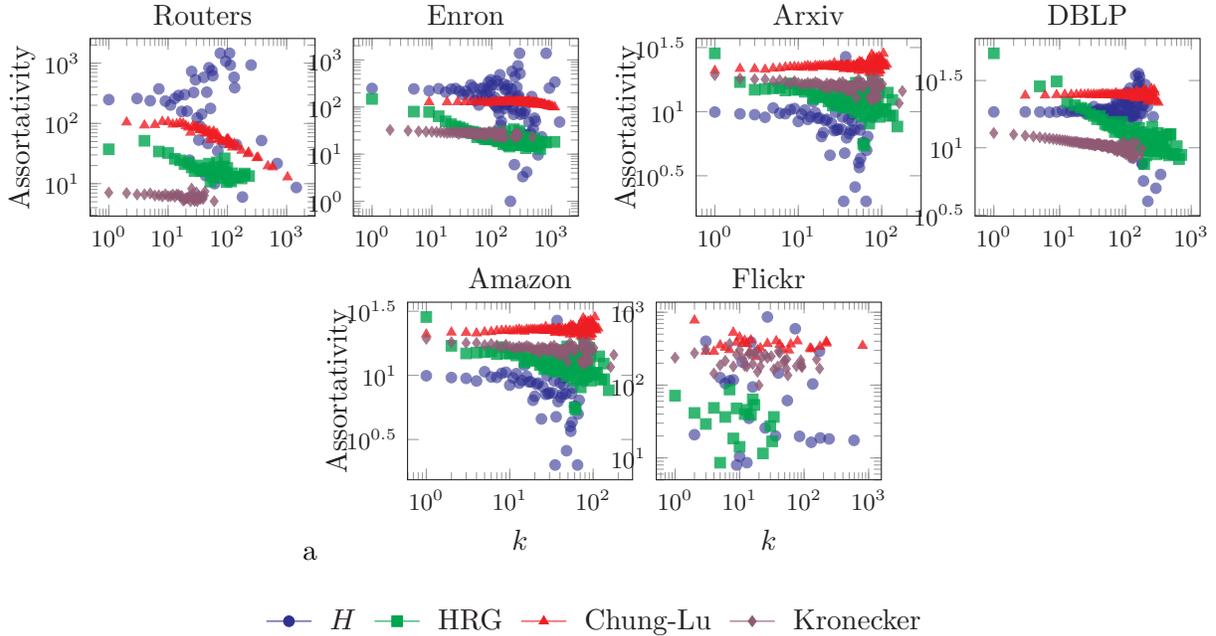

Figure 13: Local Degree Assortativity. HRG, Chung-Lu, and Kronecker graphs show mixed results with no clear winner.

**Local Degree Assortativity.** The global degree assortativity of a graph measures its tendency to have high-degree vertices connect to high-degree vertices and vice versa measured as a Pearson correlation coefficient. The local degree assortativity is measured for each vertex as the amount that each vertex contributes to the overall correlation, i.e., how different the vertex is from its neighbors. Figure 13 shows the degree assortativity for each vertex from each generated graph.

The last three graph metrics, $k$-core, local clustering coefficient, and local degree assortativity, all showed a relatively poor performance of the Chung-Lu and Kronecker graph generators. HRG modelled the $k$-core and local clustering coefficients rather well but had inconsistent results in the local degree assortativity plots.

## 5.4 Canonical Graph Comparison

The previous network properties primarily focus on statistics of the global network. However, there is mounting evidence which argues that the graphlet comparisons are a complete way to measure the similarity between two graphs [18, 17]. The graphlet distribution succinctly describes the number of small, local substructures that compose the overall graph and therefore more completely represents the details of what a graph "looks like." Furthermore, it is possible for two very dissimilar graphs to have the same degree distributions, hop plots, etc., but it is difficult for two dissimilar graphs to fool a comparison with the graphlet distribution.

Table 2: Graphlet Statistics and Graphlet Correlation Distance (GCD) for six real-world graphs. First row of each section shows the original graph's graphlet counts, remaining row shows mean counts of 10 runs for each graph generator. We find that the HRG model generates graphs that closely approximate the graphlet counts of the original graph.

| Graphs |  |  |  |  |  |  |  |  | GCD |
|---|---|---|---|---|---|---|---|---|---|
| **Routers** | 13511 | 1397413 | 9863 | 304478 | 6266541 | 177475 | 194533149 | 18615590 | |
| HRG | 13928 | 1387388 | 9997 | 288664 | 6223500 | 174787 | 208588200 | 18398430 | 1.41 |
| Kronecker | 144 | 61406 | 0 | 80 | 10676 | 973 | 642676 | 551496 | 2.81 |
| Chung-Lu | 4787 | 356897 | 6268 | 81403 | 1651445 | 13116 | 35296782 | 4992714 | 2.00 |
| **Enron** | 727044 | 23385761 | 2341639 | 22478442 | 375691411 | 6758870 | 4479591993 | 1371828020 | |
| HRG | 79131 | 4430783 | 49355 | 554240 | 13123350 | 556760 | 688165900 | 54040090 | 0.51 |
| Kronecker | 2598 | 5745412 | 1 | 1011 | 608566 | 49869 | 1.89468000 | 141065800 | 2.88 |
| Chung-Lu | 322352 | 23590260 | 1191770 | 16267140 | 342570000 | 10195620 | 3967912000 | 2170161000 | 1.33 |
| **arXiv** | 89287 | 558179 | 320385 | 635143 | 4686232 | 382032 | 11898620 | 7947374 | |
| HRG | 88108 | 606999 | 320309 | 656554 | 5200392 | 455516 | 15691941 | 9162859 | 1.10 |
| Kronecker | 436 | 224916 | 1 | 293 | 47239 | 4277 | 3280822 | 2993351 | 2.10 |
| Chung-Lu | 927 | 232276 | 6 | 967 | 87868 | 11395 | 2503333 | 3936998 | 1.82 |
| **DBLP** | 2224385 | 15107734 | 16713192 | 4764685 | 96615211 | 203394 | 258570802 | 25244735 | |
| HRG | 1271520 | 7036423 | 1809570 | 2716225 | 26536420 | 296801 | 71099374 | 28744359 | 1.59 |
| Kronecker | 869 | 21456020 | 0 | 25 | 150377 | 11568 | 517370300 | 367981700 | 2.82 |
| Chung-Lu | 1718 | 22816460 | 740 | 91 | 306993 | 27856 | 453408500 | 495492000 | 1.73 |
| **Amazon** | 5426197 | 81876562 | 4202503 | 39339842 | 306482275 | 10982173486 | 11224584 | 1511382488 | |
| HRG | 4558006 | 90882984 | 3782253 | 35405858 | 275834048 | 12519677774 | 10326617 | 1556723963 | – |
| Kronecker | 11265 | 118261600 | 40 | 1646 | 4548699 | 350162 | 6671637000 | 4752968000 | – |
| Chung-Lu | 4535 | 71288780 | 21 | 6376 | 5874750 | 95323 | 11008170000 | 2134629000 | – |
| **Flickr** | 24553 | 3754965 | 1612 | 38327 | 2547637 | 63476 | 197979760 | 30734524 | |
| HRG | 24125 | 4648108 | 1600 | 39582 | 3130621 | 68739 | 409838400 | 41498780 | – |
| Kronecker | 679294 | 494779400 | 16068 | 4503724 | 951038500 | 78799860 | 96664230000 | 76331380000 | – |
| Chung-Lu | 7059002 | 787155400 | 5003082 | 313863800 | 12826040000 | 1513807000 | 168423000000 | 247999700000 | – |

Table 2 shows the mean graphlet counts over 10 runs for each graph generator. We find that graphlet counts for the graphs generated by HRG follow the original counts more closely, and in many cases much more closely, than the Kronecker and Chung-Lu graphs.

**Graphlet Correlation Distance**  Recent work from systems biology has identified a new metric called the Graphlet Correlation Distance (GCD). The GCD computes the distance between two graphlet correlation matrices – one matrix for each graph [19]. It measures the frequency of the various graphlets present in each graph, i.e., the number of edges, wedges, triangles, squares, 4-cliques, etc., and compares the graphlet frequencies of each node across two graphs. Because the GCD is a distance metric, lower values are better. The GCD can range from $[0, +\infty]$, where the GCD is 0 if the two graphs are isomorphic.

The rightmost column in Tab. 2 shows the GCD results. Unfortunately, the node-by-node graphlet enumerator used to calculate the GCD [19] could not process the large Amazon and Flickr graphs, so only the summary graphlet counts are listed for the two larger graphs [20]. The results here are clear: HRG significantly outperforms the Chung-Lu and Kronecker models. The GCD opens a whole new line of network comparison methods that stress the graph generators in various ways. We explore many of these options next.

## 5.5 Graph Extrapolation

Recall that HRG learns the grammar from $k = 4$ subgraph-samples from the original graph. In essence, HRG is extrapolating the learned subgraphs into a full-size graph. This raises the question: if we only had access to a small subset of some larger network, could we use our models to infer a larger (or smaller) network with the same local and global properties? For example, given the 34-node Karate Club graph, could we infer what a Karate Club might look like if it's membership doubled?

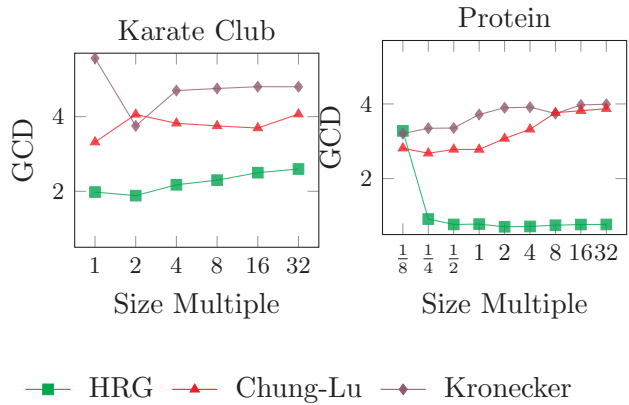

Figure 14: GCD of graphs extrapolated in multiples up to 32x from two small graphs. HRG outperforms Chung-Lu and Kronecker models when generating larger graphs. Lower is better.

Using two smaller graphs, Zachary's Karate Club (34 nodes, 78 edges) and the protein-protein interaction network of *S. cerevisiae* yeast (see Table 1), we learned an HRG model with $k = 1$ and $s = n$, i.e., no sampling, and generated networks of size-$n^* =$ 2x, 3x, ..., 32x. For the protein graph, we also sampled down to $n^* = x/8$. Powers of 2 were used because the standard Kronecker model can only generate graphs of that size. The Chung-Lu model requires a size-$n^*$ degree distribution as input. To create the proper degree distribution we fitted a Poisson distribution ($\lambda = 2.43$) and a Geometric Distribution ($p = 0.29$) to Karate and Protein graphs respectively and drew $n^*$ degree-samples from their respective distributions. In all cases, we generated 20 graphs at each size-point.

Rather than comparing raw numbers of graphlets, the GCD metric compares the *correlation* of the resulting graphlet distributions. As a result, GCD is largely immune to changes in graph size. Thus, GCD is a good metric for this extrapolation task. Figure 14 shows the mean GCD scores; not only does HRG generate good results at $n^* = 1$x, the GCD scores remain mostly level as $n^*$ grows.

## 5.6 Infinity Mirror

Next, we characterize the robustness of graph generators by introducing a new kind of test we call the *infinity mirror*.[1] One of the motivating questions behind this idea was to see if HRG holds sufficient information to be used as a reference itself. In this test, we repeatedly learn a model from a graph generated by an earlier version of the same model. For HRG, this means that we learn a set of production rules from the original graph $H$ and generate a new graph $H^*$; then we set $H \leftarrow H^*$ and repeat whereby learning a new model from the generated graph recursively. We repeat this process ten times and compare the output of the 10th recurrence with the original graph using GCD.

We expect to see that all models degenerate over ten recurrences. The question is, how quickly do the models degenerate and how badly do the graphs become?

Figure 15 shows the GCD scores for the HRG, Chung-Lu and Kronecker models at each recurrence (we have also validated the Infinity Mirror tests with other variations to the Chung-Lu model including the Block Two-Level Erdős-Rényi Model with similar results [41]). Surprisingly, we find that HRG stays steady, and even improves its performance while the Kronecker and Chung-Lu models steadily decrease their performance as expected. We do not yet know why HRG improves performance in some cases. Because GCD measures the graphlet correlations between two graphs, the improvement in GCD may be because HRG is implicitly homing in on rules that generate the necessary graph patterns.

---
[1] "Infinity mirror" gets its name from the novelty item with a pair of mirrors, set up to create a series of smaller and smaller reflections that appear to taper to an infinite distance.

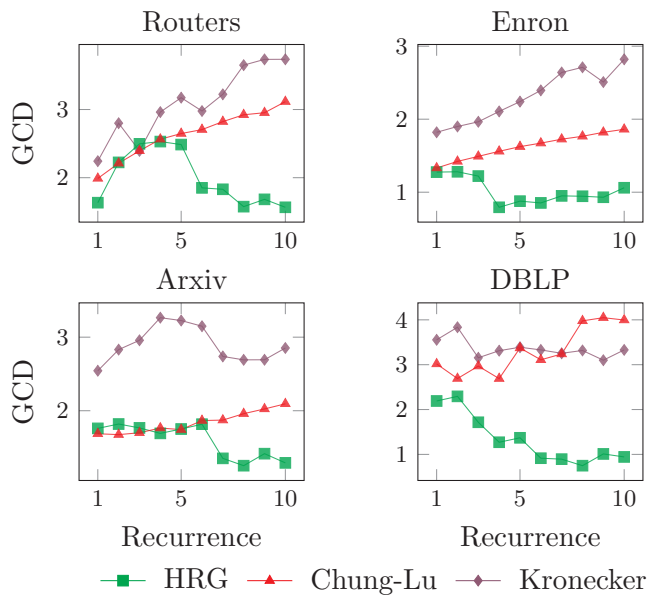

Figure 15: Infinity Mirror: GCD comparison after each recurrence. Unlike Kronecker and Chung-Lu models, HRG does not degenerate as its model is applied repeatedly.

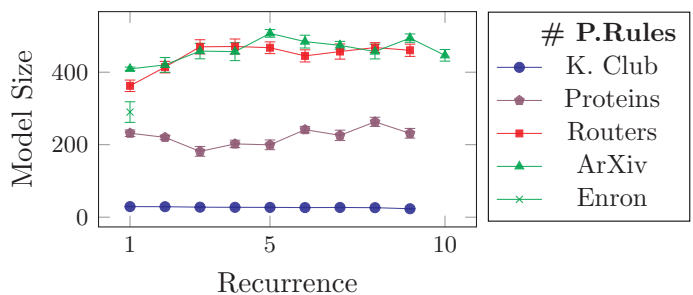

Figure 16: Number of rules (mean over 20 runs) derived as the number of recurrences increases.

### 5.6.1 Infinity Mirror Model Size

The number of production rules derived from a given graph using Fixed-Size Graph Generation. Fig. 16 shows the number of nodes in graphs after 1, 5, and 10 feedback iterations. The trend for each input graph varies slightly, but in general the model-size (i.e., the number of production rules derived) stays flat.

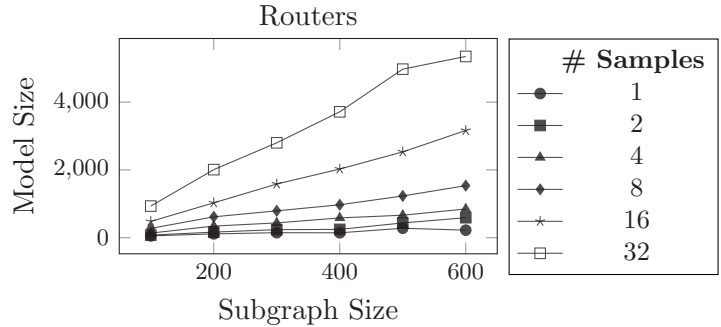

Figure 17: HRG model size as the subgraph size $s$ and the number of subgraph samples $k$ varies. The model size grows linearly with $k$ and $s$.

## 5.7 Sampling and Grammar Complexity

We have shown that HRG can generate graphs that match the original graph from $k = 4$ samples of $s = 500$-node subgraphs. If we adjust the size of the subgraph, then the size of the clique tree will change causing the grammar to change in size and complexity. A large clique tree ought to create more rules and a more complex grammar, resulting in a larger model size and better performance; while a small clique tree ought to create fewer rules and a less complex grammar, resulting in a smaller model size and a lower performance.

To test this hypothesis, we generated graphs by varying the number of subgraph samples $k$ from 1 to 32, while also varying the size of the sampled subgraph $s$ from 100 to 600 nodes. Again, we generated 20 graphs for each parameter setting. Figure 17 shows how the model size grows as the sampling procedure changes on the Internet Routers graph. Plots for other graphs show a similar growth rate and shape but are omitted due to space constraints.

To test the statistical correlation we calculated Pearson's correlation coefficient between the model size and sampling parameters. We find that the $k$ is slightly correlated with the model size on Routers ($r = 0.31$, $p = 0.07$), Enron ($r = 0.27, p = 0.09$), arXiv ($r = 0.21, p = 0.11$), and DBLP ($r = 0.29$, $p = 0.09$). Furthermore, the choice of $s$ affects the size of the clique tree from which the grammars are inferred. So its not surprising that $s$ is highly correlated with the model size on Routers ($r = 0.64$), Enron ($r = 0.71$), arXiv ($r = 0.68$), and DBLP ($r = 0.54$) all with $p \ll 0.001$.

Because we merge identical rules when possible, we suspect that the overall growth of the HRG model follows Heaps law [42], i.e., that the model size of a graph can be predicted from its rules; although we save a more thorough examination of the grammar rules as a matter for future work.

### 5.7.1 Model size and Performance

One of the disadvantages of the HRG model, as indicated in Fig. 17, is that the model size can grow to be very large. But this again begs the question: do larger and more complex HRG models result in improved performance?

To answer this question, we computed the GCD distance between the original graph and graphs generated by varying $k$ and $s$. Figure 18 illustrates the relationship between model size and the GCD. We use the Router and DBLP graphs to shows the largest and smaller of our datasets; other graphs show similar results, but we omit their plots due to space. Surprisingly, we find that the performance of models with only 100 rules is similar to the performance of the largest models. In the Router results, two very small models with poor performance had only 18 and 20 rules each. Best fit lines are drawn to illustrate the axes relationship where negative slope indicates that larger models perform better. Outliers can dramatically affect the outcome of best-fit lines, so the faint line in the Routers graph shows the best fit line if we remove the two square outlier points. Without removing outliers, we find only a slightly negative slope on the best fit line indicating only a slight performance improvement between HRG models with 100 rules and HRG models with 1,000 rules. Pearson's correlation coefficient comparing GCD and model size similarly show slightly negative correlations on Routers ($r = -0.12, p = 0.49$), Enron ($r = -0.09, p = 0.21$), ArXiv ($r = 0.04, p = 0.54$), and DBLP ($r = -0.08, p = 0.62$)

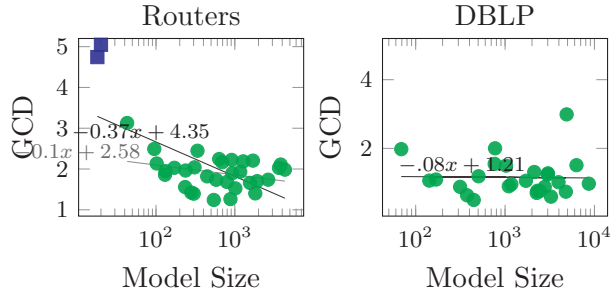

Figure 18: GCD as a function of model size. We find a slightly negative relationship between model size and performance, but with quickly diminishing returns. We show best-fit lines and their equations; the shorter fit line in the Routers plot ignores the square outlier points.

### 5.7.2 Runtime Analysis

The overall execution time of the HRG model is best viewed in two parts: (1) rule extraction, and (2) graph generation.

Unfortunately, finding a clique tree with minimal width i.e., the treewidth $tw$, is NP-Complete. Let $n$ and $m$ be the number of vertices and edges respectively in $H$. Tarjan and Yannikakis' Maximum Cardinality Search (MCS) algorithm finds usable clique trees [27] in linear time $O(n+m)$, but is not guaranteed to be minimal.

The running time of the HRG rule extraction process is determined exclusively by the size of the clique tree as well as the number of vertices in each clique tree node. From Defn. 2.1 we have that the number of nodes in the clique tree is $m$. When minimal, the number of vertices in the largest clique tree node $\max(|\eta_i|)$ (minus 1) is defined as the treewidth $tw$. However, clique trees generated by MCS have $\max(|\eta_i|)$ bounded by the maximum degree of $H$ and is denoted as $\Delta$ [43]. Therefore, given an elimination ordering from MCS, the computational complexity of the extraction process is in $O(m \cdot \Delta)$. In our experiments, we perform $k$ samples of size-$s$ subgraphs. So, when sampling with $k$ and $s$, we amend the runtime complexity to be $O(k \cdot m \cdot \Delta)$ where $m$ is bounded by the number of hyperedges in the size-$s$ subgraph sample and $\Delta \leq s$.

Graph generation requires a straightforward application of rules that is linear in the number of edges in the output graph.

We performed all experiments on a modern consumer-grade laptop in an unoptimized, unthreaded python implementation. We recorded the extraction time while generating graphs for the size-to-GCD comparison in the previous section. Although the runtime analysis gives theoretical upper bounds to the rule extraction process, Fig. 19 shows that the extraction runtime is highly correlated to the size of the model in Routers ($r = 0.68$), arXiv ($r = 0.91$), Enron ($r = 0.88$), and DBLP ($r = 0.94$) all with $p \ll 0.001$. Simply put, more rules require more time, but there are diminishing returns. So it may not be necessary to learn complex models when smaller HRG

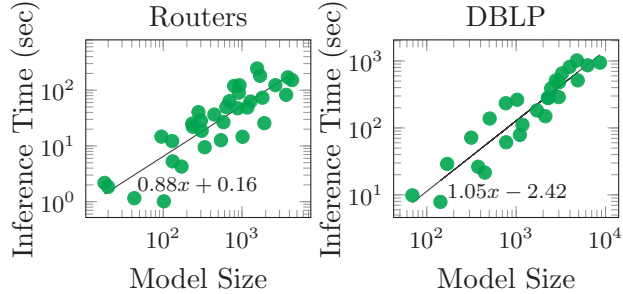

Figure 19: Total extraction runtime (i.e., clique tree creation and rule extraction) as a function of model size. Best fit lines on the log-log plot show that the execution time grows linearly with the model size.

models tend to perform reasonably well.

By comparison, the Kronecker graph generator learns a model in $O(m)$ and can create a graph in $O(m)$. The Chung-Lu model does not learn a model, but rather takes, as input, a degree sequence; graph generation is in $O(n + m)$.

### 5.7.3 Graph Guarantees

In earlier work we showed that an application of HRG rules corresponding to a traversal of the clique tree will generate an isomorphic copy of the original graph [24].

Unlike the Kronecker and Chung-Lu graph generators, which are guaranteed to generate graph with power-law degree distributions, there are no such guarantees that can be made about the shape of graphs generated by HRGs. The reason is straightforward: the HRG generator is capable of applying rules in any order, therefore, a wide variety of graphs are possible, although improbable, given an HRG grammar.

But the lack of a formal guarantees give the HRG model flexibility to model a large variety of graphs. For example, given a line-graph, the HRG model will generate a new graph that looks, more-or-less, like a line-graph. If given a random graph, characterized by a binomial degree distribution, then HRG is likely to generate a new graph with a binomial degree distribution.

## 6  Conclusions

This paper describes a new generative network framework that learns a hyperedge replacement grammar (HRG) given a simple, general graph and grows new graphs. The inference (or model learning) step uses clique trees (also known as junction trees, tree decomposition, intersection trees) to extract an HRG, which characterizes a set of production rules. We show that depending on how HRG grammar rules are applied, during the graph generation step, the resulting graph is isomorphic to the original graph if the clique tree is traversed during reconstruction. More significantly, we show that a stochastic application of the HRG grammar rules creates new graphs that have very similar properties to the original graph. The results of graphlet correlation distance experiments, extrapolation, and the infinity mirror are particularly exciting because our results show a stark improvement in performance over several existing graph generators.

Perhaps the most significant finding that comes from this work is the ability to interrogate the generation of substructures and subgraphs within the grammar rules that combine to create a holistic graph. Forward applications of the technology described in this work may allow us to identify novel patterns analogous to the previously discovered triadic closure and bridge patterns found in real-world social networks. Thus, an investigation into the nature of the extracted rules and their meaning (if any) is a top priority.

In the future, we plan to investigate differences between the grammars extracted from different types of graphs; we are also interested in exploring the implications of finding two graphs which have a large overlap in their extracted grammars. Among the many areas for future work that this study opens, we are particularly interested in learning a grammar from the actual growth of some dynamic or evolving graph. Within the computational theory community, there has been a renewed interest in quickly finding clique trees of large real-world graphs that are closer to optimal. Because of the close relationship of HRG and clique trees are shown in this paper, any advancement in clique tree algorithms could directly improve the speed and accuracy of graph generation.

We encourage the community to explore further work bringing HRGs to attributed graphs, heterogeneous graphs and developing practical applications of the extracted rules. Given the current limitation related to the growth in the number of extracted rules as well as the encouraging results from small models, we are also looking for sparsification techniques that might limit the model's size while still maintaining performance.

## Acknowledgments

This work is funded by a grant from the Templeton Foundation (FP053369-M/O) and NSF IIS (#1652492).